\documentclass[showpacs,amsmath,amssymb,prl,superscriptaddress,floatfix,twocolumn]{revtex4}
\usepackage{graphicx}
\usepackage{subfigure}
\usepackage{dcolumn}
\usepackage{bm}
\newcommand{\bs}{\boldsymbol}
\newcommand{\ket}{\rangle}
\newcommand{\bra}{\langle}
\begin{document}

\title{MIT bag model inspired partonic transverse momentum distribution for prompt photon production in pp collisions}
\author{F.K. Diakonos}
\author{N.K. Kaplis}
\author{X.N. Maintas}
\affiliation{Department of Physics, University of Athens, GR-15771 Athens, Greece}
\date{\today}

\begin{abstract}
We consider the prompt photon production in $pp$ collisions using, within the framework of perturbative QCD, a non-gaussian distribution for the transverse momentum distribution of the partons inside the proton. Our description adopts the widely used in the literature factorization of the partonic momentum distribution into longitudinal and transverse components. It is argued that the non-gaussian distribution of the intrinsic transverse momenta of the partons is dictated by the asymptotic freedom as well as the $3D$ confinement of the partons in the proton. To make this association more transparent we use the MIT bag model, which plainly incorporates both properties (asymptotic freedom, confinement), in order to determine in a simplified way the partonic transverse momentum distribution. A large set of data from 6 different experiments have been fitted with this simple description using as a single free parameter the mean partonic transverse momentum $ \langle k_T \rangle$. Surprisingly enough, a perfect fit of the experimental data turns out to require $ \langle k_T\rangle$ values which are compatible with Heisenberg's uncertainty relation for the proton and decrease almost smoothly as a function of the scaled variable $z=\frac{p_T}{\sqrt{s}}$, where $p_T$ is the transverse momentum of the final photon and $\sqrt{s}$ is the beam energy in the center of mass frame. Our analysis indicates that asymptotic freedom and $3D$ confinement may influence significantly the form of the partonic transverse momentum distribution leaving an imprint on the $pp \to \gamma + X$ cross section.
\end{abstract}

\noindent
\pacs{13.60.Le,13.85.Ni,12.38.Qk}
\maketitle

The production of photons with large transverse momentum is an excellent probe of the dynamics in hard scattering processes \cite{Owens87,Baer90}. In particular, the study of direct photon production possesses numerous and well known advantages, both theoretical and experimental
\cite{Baer90,Camilleri88,Peressounko06,Aurenche88,Aurenche90,Berger91,Gordon94,Vogelsang95}. In the latter case the main advantage is
that photons are easier  to detect than jets. From the theoretical point of view the main advantage is the simplicity of the process allowing for an accurate determination of the gluon distribution within the proton. In the lowest order (${\cal{O}}(\alpha \alpha_s)$) only two subprocesses, $g q \to \gamma q$ (Compton) and $q \bar{q} \to \gamma g$ (annihilation), contribute to high $p_T$ photons. Their characteristic signature is the production of a photon isolated from the hadrons in the event, accompanied by a kinematically balancing high-$p_T$ jet appearing on the opposite site. In the next-to-leading order (NLO) the process associated with the production of a photon coming from the collinear fragmentation of a hard parton produced in a short-distance subprocess, constitutes a background to the direct photon production of the same order in $\alpha_s$ as the corresponding Born level terms \cite{Catani02} provided that the fragmentation scale is large enough. However, the contribution from fragmentation remains small (less than $10\%$) for fixed target experiments and becomes significant only in inclusive prompt photon production at higher collider energies
\cite{Catani02}. Recently there has been observed a systematic disagreement between theoretical NLO predictions
\cite{Florian05,Aurenche84,Aurenche88,Baer90,Gordon94,Aversa89,Florian03,Jager03} and experimental data \cite{Apanasevich98,Ballocchi98} for prompt
photon production which cannot be globally improved adapting the gluon distribution function. Especially for fixed target experiments NLO approximation
shows a significant underestimation of the cross section for some of the measured data sets \cite{Florian05,Apanasevich98}. A similar discrepancy can
be observed between NLO calculations and the experimental data of inclusive single neutral pion production: $pp \to \pi^0 X$ in mostly the same experiments as in the photon case \cite{Florian0501}. For the pion production the theoretical description is improved by taking into account certain large contributions to the partonic hard scattering cross section to all orders in perturbation theory using the technique of threshold resummation
\cite{Florian0501,Sterman87}. The same technique can be applied to the photon production by calculating the QCD resummation contribution to the
partonic processes $qg \to \gamma q$ and $q \bar{q} \to \gamma g$ \cite{Laenen98,Catani99}. However, the result is a relatively small enhancement,
not enough to compensate for the gap between the prompt photon data in fixed target experiments \cite{Florian05,Catani99} and theoretical predictions. An additional improvement can be achieved by including in the theoretical treatment resummation effect to the fragmentation component succeeding in this way a good description of UA6 and R806 $pp$ data but still failing to reproduce the data of E706 \cite{Florian05}. The conclusion of this analysis is that resummed theoretical results present a residual shortfall in the description of photon and pion production data in fixed target experiments. One possible explanation of this effect is the existence of a non-perturbative contribution associated with intrinsic partonic transverse momentum $k_T$ \cite{Peressounko06,Aurenche06,Florian05}. To incorporate this effect in the conventional pQCD one assumes a factorization ansatz based on the statistical independence between longitudinal and transverse momenta of the partons. In this treatment the distribution of intrinsic transverse momentum $g(k_T)$ is taken to have a gaussian form as suggested by the early work of Dalitz \cite{Dalitz65}. Using the so called $k_T$-smearing one can
fit most of the experimental data \cite{Sivers76,Owens87,Wang97}. However the method suffers from two serious disadvantages concerning the values of the
introduced non-perturbative parameter $\langle k_T \rangle$ needed to fit the data: (a) they are incompatible with the estimation of the proton transverse radius according to Heisenberg's uncertainty relation and (b) they do not depend smoothly on physical parameters of the process as beam energy or the transverse momentum of the produced particle. These shortcomings remain even with the inclusion of higher order contributions in perturbation theory making the $k_T$-smearing approach unattractive from the theoretical point of view.

Within the framework of pQCD an alternative scenario for the description of $pp$ collisions has been recently proposed \cite{Diakonos06}, incorporating the calculation of the partonic transverse momentum distribution using a phenomenological quark potential model introduced in the past to describe baryonic spectra \cite{Martin81,Richard81}. The underlying idea in this treatment is to determine the influence of bound state effects on the shape of $g(k_T)$. In the calculation performed in \cite{Diakonos06} $g(k_T)$ turns out to have a non-gaussian profile with a characteristic tail. Furthermore, when applied to the description of the $pp \to \pi^0 + X$ process, one succeeds to overcome the disadvantages of the gaussian $g(k_T)$ discussed above. However, there are still unsatisfactory issues in this treatment since a connection with first principles is missing and the asymptotic freedom property in the partonic dynamics is not taken into account.

In the present work we will argue that a non-gaussian transverse momentum distribution with a slowly vanishing tail, may originate from the fundamental properties of asymptotic freedom and $3D$ confinement of the partonic degrees of freedom in the proton. In addition, using the direct photon production in $pp$ collisions, it will be revealed that, within the proposed scenario, the values of $\langle k_T \rangle$, needed for the description of all the available experimental data, are characterized by a remarkable smooth dependence on the scaled variable $z=\frac{p_T}{\sqrt{s}}$ where $p_T$ is the transverse momentum of the produced photon and $\sqrt{s}$ the beam energy in the center of mass frame. The results of our analysis clearly support the initial assumption that part of the residual shortfall in the perturbative description of $\gamma$ or $\pi$ production in $pp$ collisions is associated with the {\it transversal confinement} of the partons in the protons.

In the next to leading order (NLO) of perturbation theory the differential cross section of the single photon production in $pp$ collision can be written as:
\begin{eqnarray}
E_{\gamma} \frac{d^3 \sigma}{d^3 p}(pp \rightarrow \gamma + X)=
K(p_T,\sqrt{s}) \sum_{abc} \int dx_a dx_b \cdot\nonumber \\
\cdot f_{a/p}(x_a,Q^2) f_{b/p}(x_b,Q^2)\cdot \frac{\hat{s}}{\pi}\frac{d \sigma}{d
\hat{t}}(ab \rightarrow c \gamma) \delta(\hat{s}+\hat{t}+\hat{u})
\label{eq:eq1}
\end{eqnarray}
where $f_{i/p}$ ($i=a,b$) are the NLO longitudinal parton distribution functions (PDF) for the colliding partons $a$ and $b$ as a function of longitudinal momentum fraction $x_i$ and factorization scale $Q$ \cite{MRST}. $\frac{d \sigma}{d \hat{t}}$ is the cross section for the partonic
subprocesses as a function of the Mandelstam variables $\hat{s},~\hat{t},~\hat{u}$ \cite{Owens87}. The higher order corrections in the
partonic subprocesses are effectively included in (\ref{eq:eq1}) through the $K$-factor, appearing in the right hand side, which depends on the transverse momentum of the outcoming photon and the beam energy \cite{Barnafoldi01}. According to our treatment we first attempt to describe
experimental data using a minimal modification of the NLO pQCD introducing partonic transverse degrees of freedom through the replacement
\cite{Owens87,Aurenche06}:
\begin{equation}
dx_i~f_{i/p}(x_i,Q^2) \longrightarrow dx_i d^2 k_{T,i} g(\bs{k}_{T,i}) f_{i/p}(x_i,Q^2)
\label{eq:eq2}
\end{equation}
in the PDF of the colliding partons ($i=a,b$). To avoid singularities in the partonic subprocesses we introduce a regularizing parton mass
\cite{Feynman78,Wang97} with value close to the constituent quark mass $m_q=0.3~GeV$ in the Mandelstam variables occurring in the denominator of the corresponding matrix elements \footnote{In fact $m$ can be chosen in the range $[0.1,1.0]~GeV$ without affecting the following analysis}. Contrary to the usual treatment where $g(\vec{k}_T)$ is taken to be a gaussian, here we will use a different form inspired by the MIT bag model \cite{MITbag}. The main advantage of this model is that it incorporates the basic features of strongly interacting quark matter, namely, asymptotic freedom and $3D$ confinement in a very simple fashion. Despite this simplicity, a self-consistent approach to obtain partonic momentum distributions within the framework of the MIT bag model is techniqally very complicated \cite{Schreiber91} leading to unintegrated PDF which are in general incompatible with the factorization ansatz (\ref{eq:eq2}). In addition it is not clear how to embed correctly such distribution functions in the pQCD scheme (\ref{eq:eq1}). However, since our approach is purely phenomenological, one can use the MIT bag model wave function for the ground state of a parton inside the proton:
\begin{equation}
\Psi(\vec{r})=N\cdot\left( \begin{array}{c}
 j_0(E\cdot r)\cdot Y_{00}\left(\begin{array}{c} 1\\0 \end{array}\right)\\
-{\imath \over \sqrt{3}} j_1 (E\cdot r)\left(\begin{array}{c}
-Y_{10}\\\sqrt{2}Y_{11} \end{array}\right)
\end{array}
 \right)
\label{eq:eq3}
\end{equation}
to obtain the partonic wave function $\phi(\vert \vec{k} \vert)$ in momentum space through the Fourier transformation:
\begin{eqnarray}
\phi(\vert \vec{k} \vert) &=& \int d^3 \bs{r} \exp[\imath\bs{k}\cdot \bs{r}] \Psi(\bs{r}) \nonumber \\
&=& \int d\cos\theta d\phi r^2dr \exp[\imath k r \cos\theta]\Psi(\bs{r})\nonumber\\
&=&\left( \begin{array}{cccc} 2\sqrt{\pi} \Phi_0(k) \\ 0 \\ -2\imath \sqrt{\pi}\Phi_1(k)\\0 \end{array} \right) , \\
\Phi_0(k) &:=& \frac{1}{k^2}\int_{0}^{R}dr j_0(Er)kr \sin kr  \nonumber \\
\Phi_1(k) &:=& \frac{1}{k^2}\int_{0}^{R}dr j_1(Er)(kr \cos kr -\sin kr) \nonumber
\label{eq:eq4}
\end{eqnarray}
In (\ref{eq:eq3}) $N$ is a normalization constant, $j_0$, $j_1$ are spherical Bessel functions, $Y_{00}$, $Y_{10}$, $Y_{11}$ are spherical harmonics and $E$ is the energy of the parton's ground state, while in (\ref{eq:eq4}) $R$ is just the radius parameter of the MIT bag model. From $\phi(\vert \bs{k} \vert)$ it is straightforward to obtain a rough estimation of the partonic transverse momentum distribution through the projection:
\begin{equation}
 \tilde{g}(\vert \bs{k}_T \vert)=2\pi k_T\int_{-\infty}^{+\infty}dk_z
\phi^\dag(\sqrt{|\bs{k}_T|^2 + k_z^2})\phi(\sqrt{|\bs{k}_T|^2 + k_z^2})
\label{eq:eq5}
\end{equation}
The $\bra k_T \ket$ is then promoted to a free parameter by rescaling of the $k_T$ axes, introducing the distribution:
\begin{equation}
  g(k_T,\bra k_T \ket ):= \left( \frac{\bra \tilde{k}_T \ket}{\bra k_T\ket}\right)^2 \tilde{g}(k_T \cdot \frac{\bra \tilde{k}_T \ket}{\bra k_T\ket})
  \label{eq:eq6}
\end{equation}
with $\bra \tilde{k}_T \ket$ being the mean transverse momentum corresponding to $\tilde{g}(\bs{k}_T)$ of equation (\ref{eq:eq5}) and $\bra k_T \ket$ being the fitting parameter and the mean transverse momentum of $g$.
The form of the calculated distribution is presented in Fig.~1 for a particular choice of $\bra k_T \ket$.
\begin{figure}[htbp]
\centerline{\includegraphics[width=9. cm]{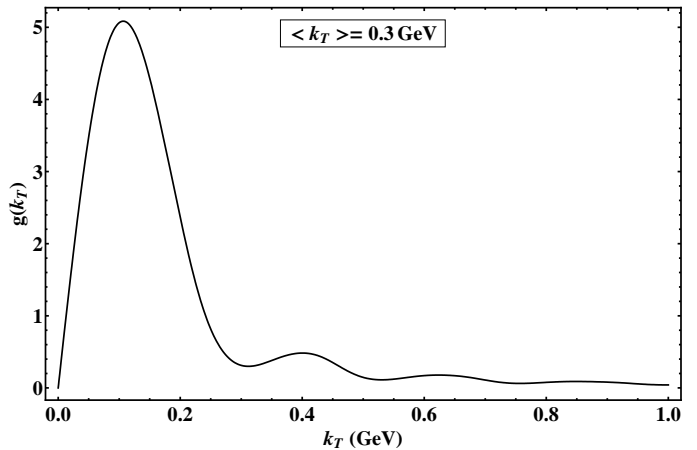}}
\caption{The partonic transverse momentum distribution $g(k_T)$ obtained using the MIT bag model.}
\label{fig:fig1}
\end{figure}
At this point it should be noticed that the slowly vanishing tail of the distribution (\ref{eq:eq5}), which turns out to be crucial for the successful description of the experimental data, originates from the $3D$ character of the problem, since the involved Bessel functions are eigenfunctions of the $3D$ Laplacian, describing the motion of the partons inside the bag.


Using eqs.(\ref{eq:eq1},\ref{eq:eq2}) we describe the data of 6 different experiments using as a single fitting parameter the $\langle k_T \rangle$-value. Within the MIT bag model such a variation can be justified as a dependence of the bag radius on $p_T$ and $\sqrt{s}$. In particular we investigate the single photon production data of the 3 CERN experiments R807 (ISR), UA6 (SPS), NA24 (SPS), as well as the FNAL experiments E704, E706 (Tevatron) and the PHENIX experiment at RHIC BNL \cite{singlegammaexp}. In Fig.~2 we summarize all the considered experimental data, shown with symbols, appropriately scaled to improve the presentation. The dashed lines represent the theoretical pQCD predictions using the resummation technique according to \cite{OwensRodos06}. The best fit results using the improved pQCD with the non-gaussian intrinsic transverse momentum distribution (Fig.~1) are presented by crosses which are not easily distinguishable since they coincide with the experimental data. In all the performed calculations we use $Q=\frac{p_T}{2}$.
\begin{figure}[htbp]
\centerline{\includegraphics[width=9. cm,height= 8. cm]{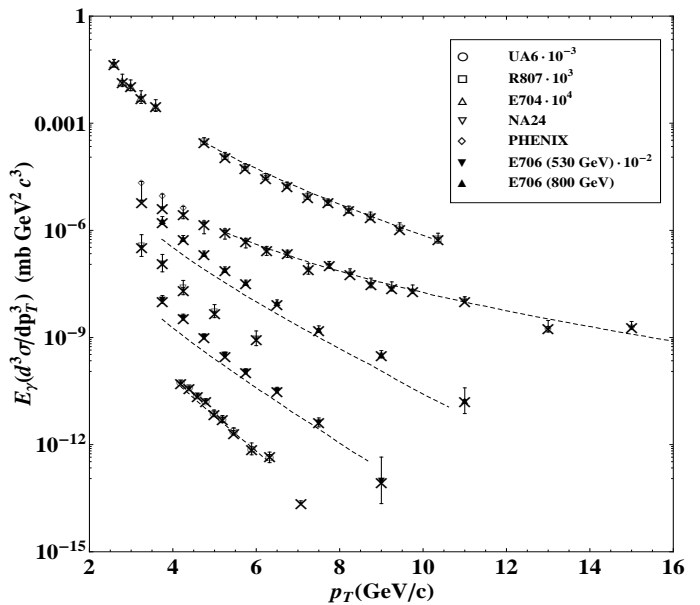}}
\caption{The cross section of prompt photon production for six different experiments at various energies. The symbols indicate the experimental data,
while the dashed line describes the theoretical results using resummation technique. The crosses indicate the fit results using the non-gaussian transverse momentum distribution of Fig.~1. The are not easily distinguishable since they coincide with the symbols presenting the experimental data.}
\label{fig:fig2}
\end{figure}
It is clearly seen that with a suitable choice of $\langle k_T \rangle$ an excellent description of the data is possible.

The significant improvement of this non-gaussian treatment is that the optimization procedure leads to $\langle k_T \rangle$-values which form a well
defined, smooth, decreasing function of the scaled parameter $z=\frac{p_T}{\sqrt{s}}$. This property is illustrated in Fig.~3 where the best fit values of $\langle k_T \rangle$, for the single gamma production in the experiments studied using a non-gaussian $k_T$-smearing (Fig.~3b), are compared with the corresponding results obtained using a gaussian distribution (Fig.~3a). It is observed that in the non-gaussian case, for $z \to 0$, the resulting $\langle k_T \rangle$-value is compatible with the transverse size of the proton according to Heisenberg's uncertainty relation. This is in fact expected since in this limit one enters into the non-perturbative regime tracing geometrical characteristics of the proton through the considered process. In addition, for increasing $z$, a decrease in the contribution of the partonic transverse momenta is anticipated, in full agreement with the non-gaussian result, since in this region the perturbative treatment becomes more and more accurate. In contrary, as it can be seen in Fig.~3a, the gaussian ansatz leads to absurd behavior incompatible with physical intuition.

\begin{figure}
\centering
\subfigure[~gaussian] 
{
    \label{fig:sub:a}
    \includegraphics[width=9cm]{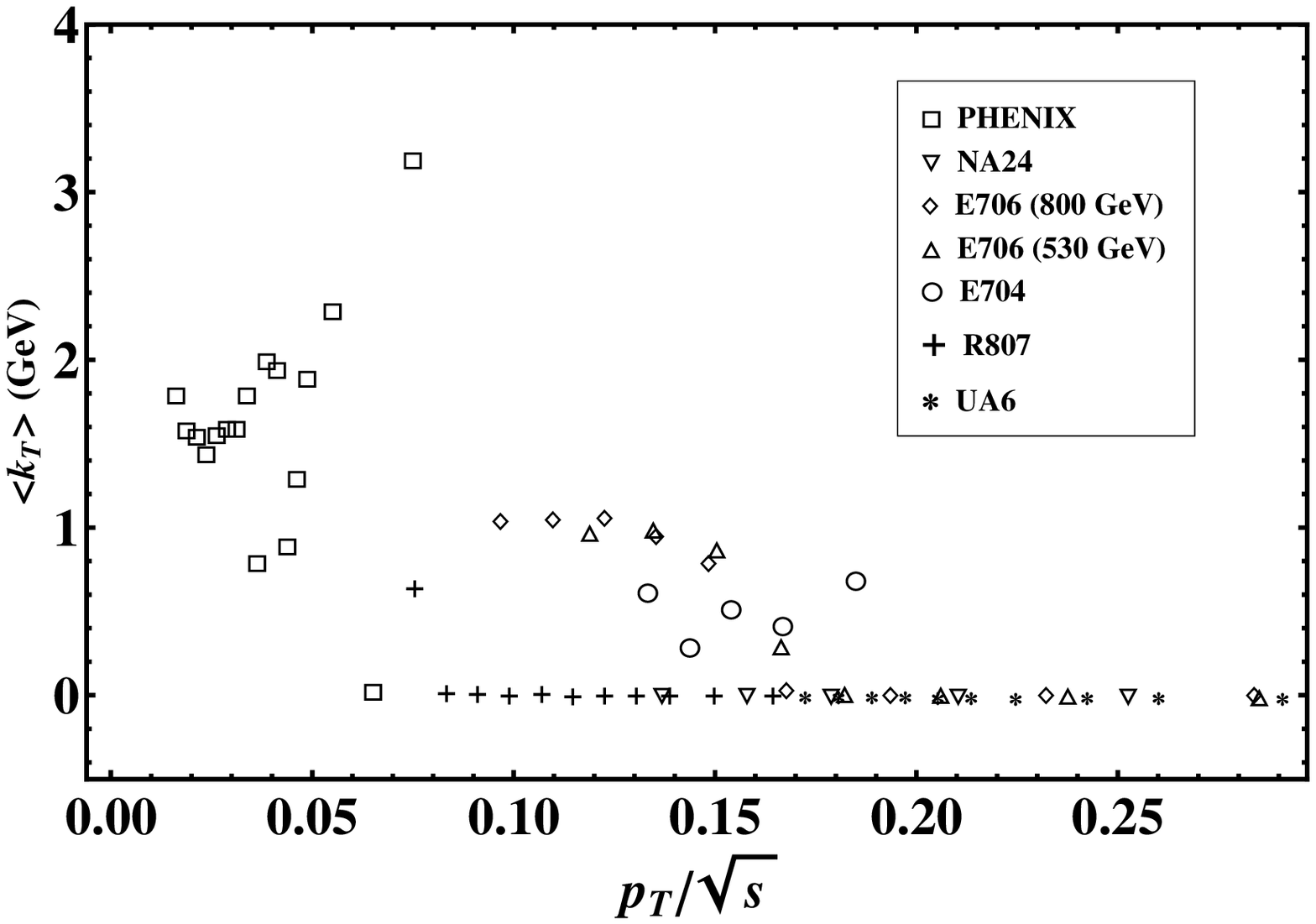}
}
\hspace{1cm}
\subfigure[~non-gaussian] 
{
    \label{fig:sub:b}
    \includegraphics[width=9cm]{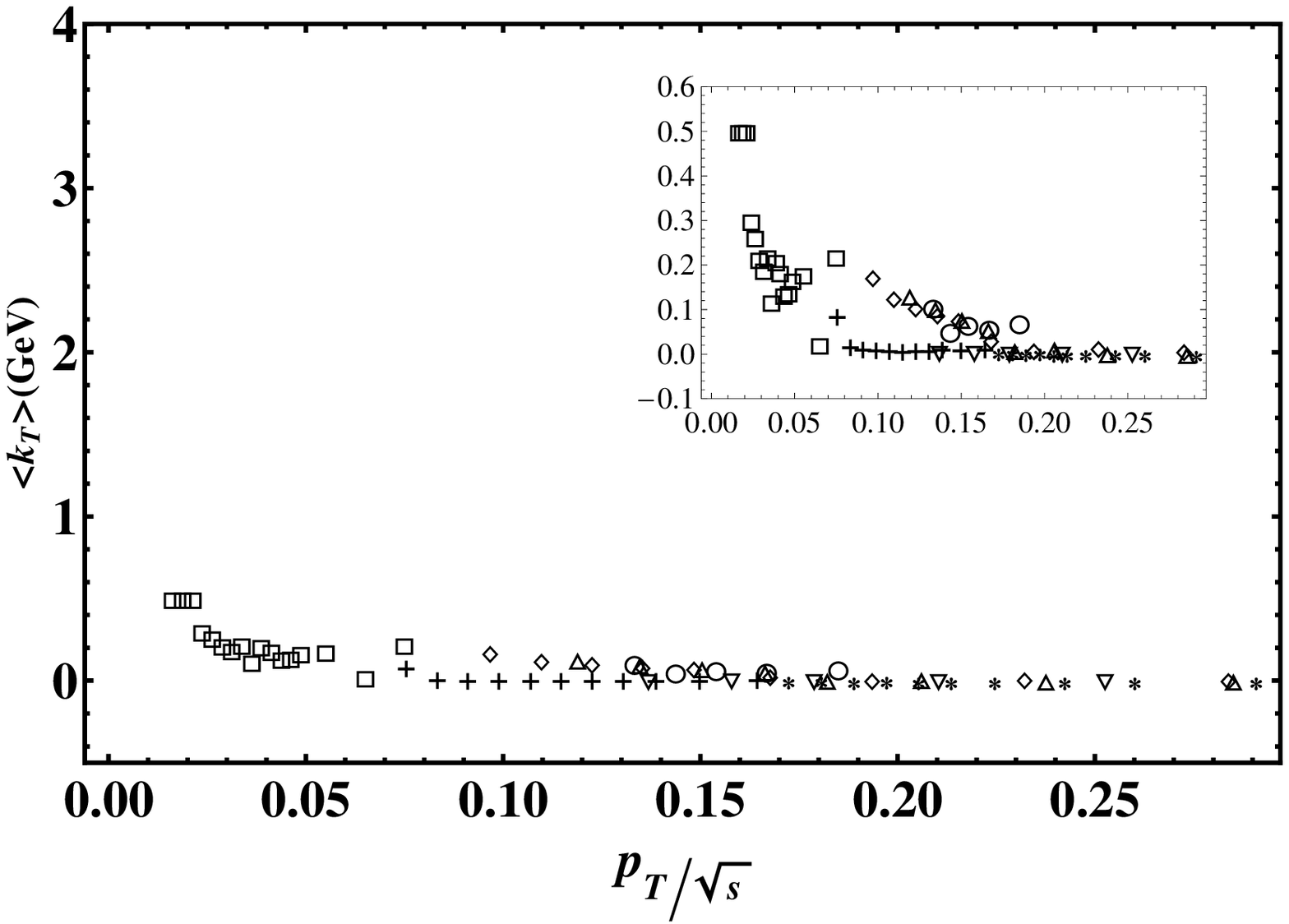}
}
\caption{The $\langle k_T \rangle$ used in the description of six different experimental datasets for single photon production, referred in the text,
(a) with a gaussian and (b) with a non-gaussian intrinsic transverse momentum distribution. The inset in Fig.~3b shows the dependence of $\langle k_T \rangle$ on $z$ in a refined scale. }
\label{fig:sub} 
\end{figure}

At this point it should be noticed that analogous calculation for $\pi_0$ production in $pp$ collisions, have been performed leading to $\langle k_T
\rangle$ values with similar characteristics. In a more complete treatment one should also calculate the cross sections for DIS, Drell-Yan and $p\bar{p}$ processes using $\langle k_T \rangle$-values depending on $z$ as dictated by the analysis of the $\gamma$ production.

In conclusion we have proposed a pQCD scheme for the description of the existing experimental data of prompt gamma production in $pp$ collisions using a non-gaussian transverse momentum distribution for the partons inside the proton inspired by the MIT bag model. The mean transverse momentum of the partons is used as a free fitting parameter. It turns out that a perfect fit of all existing data is achieved using $\langle k_T \rangle$-values which tend to become a smooth function of $z=\frac{p_T}{\sqrt{s}}$, where $p_T$ is the transverse momentum of the outgoing photon and $\sqrt{s}$ is the corresponding beam energy. In addition the range of the required $\langle k_T \rangle$-values is in accordance with the geometrical characteristics of the proton when non-perturbative effects are expected to be visible ($z \to 0$). This surprisingly successfull result suggests that the used partonic transverse momentum distribution captures essential characteristics (asymptotic freedom, $3D$ confinement) of the parton dynamics inside the proton. Thus, it looks quite promising the effort to put this purely phenomenological treatment on a more strict basis within the framework of pQCD as well as to extent it in order to describe $pA$ and $AA$ collisions by taking into account confinement characteristics of the involved partons. However these challenging questions are left for a future work.

\end{document}